# Road to Room-Temperature Superconductivity: $T_c$ above 260 K in Lanthanum Superhydride under Pressure

Russell J. Hemley[1,2], Muhtar Ahart[1], Hanyu Liu[3], and Maddury Somayazulu[4]

[1]*Department of Physics, University of Illinois at Chicago, Chicago IL 60607, USA*
[2]*Department of Chemistry, University of Illinois at Chicago, Chicago IL 60607, USA*
[3]*College of Physics, Jilin University, Changchun 130012, China*
[4]*HPCAT, X-ray Science Division, Argonne National Laboratory, Argonne IL 60439, USA*

The use of high pressure to realize superconductivity in the vicinity of room temperature has a long history, much of it focused on achieving this in hydrogen-rich materials. This paper provides a brief overview of the work presented at this May 2018 conference, together with background on motivation and techniques, the theoretical predictions of superconductivity in lanthanum hydride, and the subsequent experimental confirmation. Theoretical calculations using density-functional based structure-search methods combined with BCS-type models predicted a new class of dense, hydrogen-rich materials – superhydrides ($MH_x$, with x > 6 and M selected rare earth elements) – with superconducting critical temperatures ($T_c$) in the vicinity of room-temperature at and above 200 GPa pressures. The existence of a series of these phases in the La-H system was subsequently confirmed experimentally, and techniques were developed for their syntheses and characterization, including measurements of structural and transport properties, at megabar pressures. Four-probe electrical transport measurements of a cubic phase identified as $LaH_{10}$ display signatures of superconductivity at temperatures above 260 K near 200 GPa. The results are supported by pseudo-four probe conductivity measurements, critical current determinations, low-temperature x-ray diffraction, and magnetic susceptibility measurements. The measured high $T_c$ is in excellent agreement with the original calculations. The experiments also reveal additional superconducting phases with $T_c$ between 150 K and above 260 K. This effort highlights the novel physics in hydrogen-rich materials at high densities, the success of 'materials by design' in the discovery and creation of new materials, and the possibility of new classes of superconductors $T_c$'s at and above room temperature.

**Introduction**

High pressure and superconductivity have a long history – almost as long as the history of superconductivity itself. Some 14 years after the discovery of superconductivity, the first measurements were carried out on Sn and Sb.[1] High pressure studies of superconductors reached the megabar pressure range in the mid 1990s. These were done with direct measurements of electrical conductivity as well as, and in combination with, magnetic susceptibility techniques.[2] These studies have been used to create new superconductors from the elements, of which some 23 have been converted to date from non-superconducting to superconducting states under pressure.[2] The techniques also have been used to tune $T_c$, revealing changes in electronic structure from shifts in $T_c$ in known superconductors. Indeed, the highest $T_c$ for many years was observed in $HgBa_2Ca_2Cu_3O_{8-\delta}$ ($T_c$ = 164 K) near 30 GPa.[3,4]

Some 50 years ago, Ashcroft[5] predicted that the high-pressure atomic metallic phase calculated by Wigner and Huntington[6] to be stable under pressure (> 25 GPa) could also be a very high temperature superconductor. This conjecture was based on Bardeen-Cooper-Schrieffer considerations, given the high Debye temperature, large electron-phonon coupling, and large density of states at the Fermi level expected for the atomic alkali-like metal system that was envisaged.[5] We now know that hydrogen does not transform directly from a simple insulating molecular phase to an atomic metallic solid, but in fact passes through a series of transformations to semiconducting and semimetallic molecular phases with increasing pressure to several hundred gigapascals,[7-10] with the transition to an atomic metallic phase currently constrained to be near 500 GPa.[11] Recent calculations predict that this crystalline phase could be a very high-temperature superconductor, with a $T_c$ above 400 K.[11] On the other hand, other calculations predict that hydrogen could be a superconducting superfluid.[12] These fundamental questions cannot yet be addressed experimentally as the pressures required and the diagnostics needed to probe the superconducting state are beyond the range of current techniques.

Alternative approaches have been sought to explore the potentially interesting physics of dense metallic hydrogen, including its possible very high $T_c$ superconductivity. In another seminal paper, Carlsson and Ashcroft[13] suggested different routes to effectively reduce the transition to atomic metallic hydrogen, including the use of dopants in the structure. This prediction prompted the search for such compounds and alloys experimentally, including studies of oxygen-, carbon-, sulfur-, and rare gas-bearing materials mixed with hydrogen. These considerations, for example, motivated studies of the $H_2O$-$H_2$[14] and $CH_4$-$H_2$[15] systems, as well as later work on $H_2S$-$H_2$[16] and $Xe$-$H_2$[17]. These studies led to the discovery of interesting chemistry, but no dissociation of the molecular hydrogen was observed in these structures at the pressures explored in these experiments.

Ashcroft[18] later extended and recast the above considerations in terms of 'chemical pre-compression', a proposal in which $H_2$ molecules in dense structures might be expected to dissociate at pressures well below those required for pure hydrogen. Interest in this possibility in producing analogs of atomic metallic hydrogen was stepped up with the application of new computational structure-search methods, which began to identify new structures with very high predicted critical temperatures in the higher hydrides, starting with the important paper that predicted a $T_c$ of 235 K in $CaH_6$.[19] This work showed that increasing the atomic hydrogen content in such structures could provide close models to atomic metallic hydrogen, including the very high $T_c$ superconductivity predicted for the material.

These considerations led to an explosion of theoretical predictions of potential hydrogen-rich high $T_c$ superconductors under pressure. Indeed, the important breakthrough has been the confirmation of very high $T_c$ superconductivity in the H-S system first predicted at $T_c$ = 80-200 K.[20,21] The high critical temperatures were subsequently confirmed with the $T_c$ of 203 K at 155 GPa reported in the phase now identified as $H_3S$,[22] a result that in turn has led to numerous subsequent theoretical calculations. Although no confirmation from an independent experimental group has apparently been reported to date for this result in the H-S system, high $T_c$ has been documented for in $H_3Se$ at megabar pressures (e.g., $T_c$ = 105 K at 135 GPa).[23]

**Predictions for Rare-Earth Superhydrides**

We undertook our own calculations for possible high $T_c$ hydrides under pressure. We found that the rare earth hydride systems showed promising results theoretically, and thus we began our own structure search calculations. The enthalpies were calculated using DFT and structure search methods as implemented in the CALYPSO code.[24,25] Convex hulls were calculated for the La-H and Y-H system and predicted the stability of very hydrogen-rich compounds, some with altogether new structure types and all consisting of fully dissociated hydrogen. Those with stoichiometries of greater than 6 hydrogens per heavy atom we termed 'superhydrides'.[26] Similar results were reported by Peng et al.,[27] who also examined additional rare earth hydrides at these pressures.

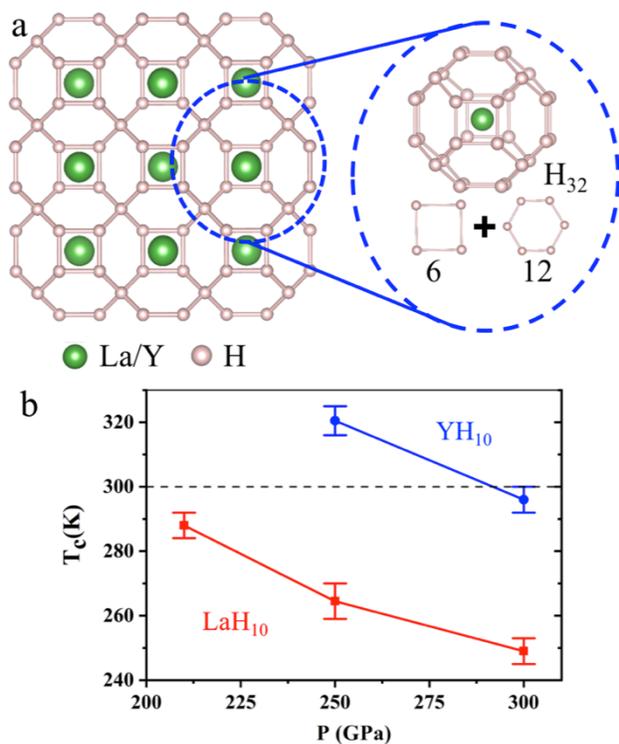

One structure that proved particularly interesting is that found theoretically for the $LaH_{10}$ and $YH_{10}$ stoichiometry (Fig. 1). This unique cubic (*Fm3m*) clathrate-type structure consists of a 32-atom cage surrounding the rare earth atom (compared to the 24-atom cage for $CaH_6$). Each cage is linked by an intriguing 8-hydrogen cube to form a highly symmetric structure.[26] Notably, the calculated H-H distances are close to those predicted for atomic metallic hydrogen, specifically in Cs(IV) structure at the similar pressures (1.1-1.2 Å).[28] The distances are appreciably longer than for molecular $H_2$, both as an isolated molecule (0.74 Å) and on compression in its dense molecular phases at megabar pressures based on spectroscopic constraints.[7] In both cubic $LaH_{10}$ and Cs(IV)-type hydrogen, the H 1*s* orbitals contribute much more than the 2*p* orbitals to the electronic density of states (DOS), and each H has 4 nearest neighbors in both structures.

**Figure 1**. (a) Predicted structure of $LaH_{10}$ and $YH_{10}$ showing the $H_{32}$ clathrate cage including its bonding components. (b) Calculated $T_c$ for $LaH_{10}$ and $YH_{10}$ as a function of pressure (after Ref. [31]).

The calculations indicated flat bands near the Fermi level giving rise to a peak in the DOS. Further, Eliashberg spectral function calculations [$\alpha^2 F(\omega)$] give a very large electron-phonon coupling parameter $\lambda$ and $T_c$ with different assumptions for a typical $\mu^*$ of 0.1-0.13 as used in previous studies. Notably, the modes soften at zone edge points and show strong electron-phonon coupling. The predicted critical temperatures have a negative pressure shift; for $LaH_{10}$ $T_c$ is 274-286 K and is highest at an instability at 210 GPa, with $\lambda = 3.41$ at this level of calculation.

**Synthesis of Lanthanum Superhydrides**

Experiments proceeded by laser-heating a mixture of La + $H_2$ in a diamond anvil cell[29] at close to the pressures of the synthesis predicted theoretically while monitoring the system with *in situ* x-ray diffraction. Major changes observed in the diffraction patterns were observed in repeated runs on heating at 170-180 GPa and 1000-1200 K and indicated a transition to an fcc-based La structure. The hydrogen stoichiometry was constrained by experimental volume systematics (i.e., independent of theory). The effective volume per hydrogen as a function of pressure for different hydrides (i.e., $AlH_3$, $IrH_3$, $FeH_5$, $FeH_3$) and the atomic volume of pure hydrogen[30] were compared to La-H with different assumed stoichiometries. LaH$_{10}$ provided the closest match based on the experimental equation of state of hydrogen (LaH$_{10\pm x}$ with x < 1, hereafter denoted LaH$_{10}$). The good agreement with the theoretically predicted equation of state of LaH$_{10}$ provided additional support to the assigned structure and stoichiometry (Fig. 2).[26]

On decompression, we found multiple phases, including structures and transitions not originally predicted.[29] This was first evident from a splitting of the cubic diffraction peaks indicative of a displacive transition to a rhombohedral structure with respect to the La sublattice. Subsequent calculations performed to examine this behavior showed this to be a soft-mode driven transition.[29] Further, optimizing the hydrogen sublattice gives a monoclinic structure (*C2/m*) in this lower pressure phase. Calculations of the electron-phonon coupling predict a $T_c$ of 220 K in this lower symmetry phase.[28]

We also investigated the nuclear quantum effects in the material using path integral molecular dynamics (PIMD) simulations. The motivation was to determine the extent to which the quantum character of this hydrogen-rich system affected the predicted stability of the cubic phase. The simulations showed that inclusion of quantum effects stabilized the cubic phase to lower pressures compared to the classical quasiharmonic calculations, in agreement with experiment.[26] Further, the PIMD calculations showed that the H-lattice is highly quantum mechanical; for example, the root-mean-squared (rms) H displacements are 20% of H-H distances at room temperature, which is close to the classical Lindemann criterion for melting. Thus, we have an interesting situation in which the predicted superconductivity persists into a temperature regime where the system begins to exhibit classical sublattice melting (Fig. 3).

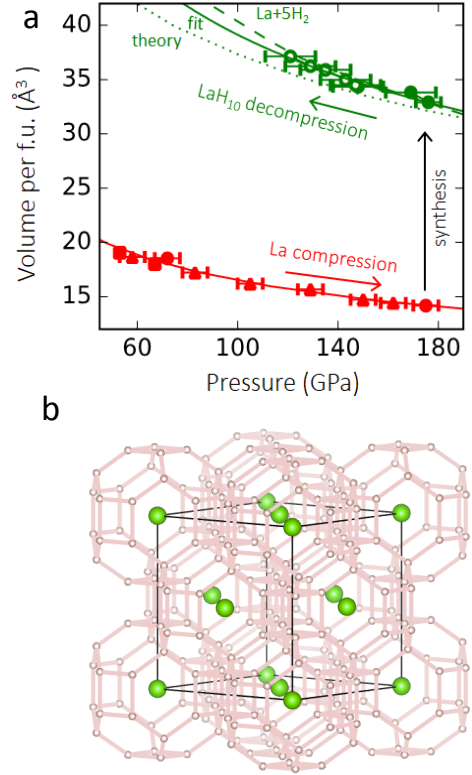

**Figure 2**. (a) Pressure–volume relations at 300 K for La (red) and LaH$_{10}$ (green) after synthesis upon laser heating; see Ref. [29] for details.

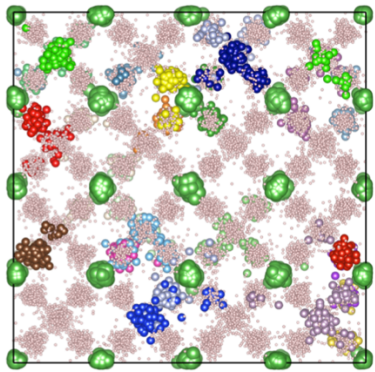

**Figure 3**. PIMD simulation of sublattice melting in cubic LaH$_{10}$ at 150 GPa. The beads show the trajectories of La (green spheres) and H (colored smaller spheres) atoms at 800 K, which is the predicted temperature of full sublattice melting.[28]

**Observations of High $T_c$ Superconductivity in LaH$_{10}$**

Having synthesized different phases in the La-H system, including the predicted LaH$_{10}$ compound, we proceeded to test the predicted high temperature superconductivity. These challenging experiments required synthesis of the material at a very high pressure together with measurements of electrical and magnetic properties of the superconducting state. We chose electrical conductivity to start, using both gaseous hydrogen and NH$_3$BH$_3$ as the hydrogen source. Following laser heating as in the earlier experiments,[29] we compared the volumes per La before and after conversion from La to $LaH_{10}$. From that very same sample with the four probes intact, we found on cooling after initial synthesis at 188 GPa, a striking drop in resistance beginning around 260 K (Fig. 4). The resistance recovers on warming, though there is a pressure shift (actually due a partial diamond anvil failure on cycling).[31]

Additional measurements using pseudo-four probe techniques confirmed these high $T_c$ results and further showed the conductivity onset varies with synthesis conditions and pressure.[31] The experiments revealed reproducible $T_c$ onsets above 260 K, in some cases as high as 280 K. Powder x-ray diffraction of $LaH_{10}$ as a function of temperature was performed to examine whether the change is associated with a structural transition. The data indicated no anomalous volume changes in the sample in this temperature range. Further experiments were done with the synthesis at higher maximum pressures, which showed that the conductivity onset occurred at higher $T_c$ with higher synthesis pressure. We suggest that this higher value is due to higher hydrogen contents resulting from the higher synthesis pressures. Current-voltage measurements obtained in these experiments (Fig. 5), provide a constraint on the critical current for the material.[31]

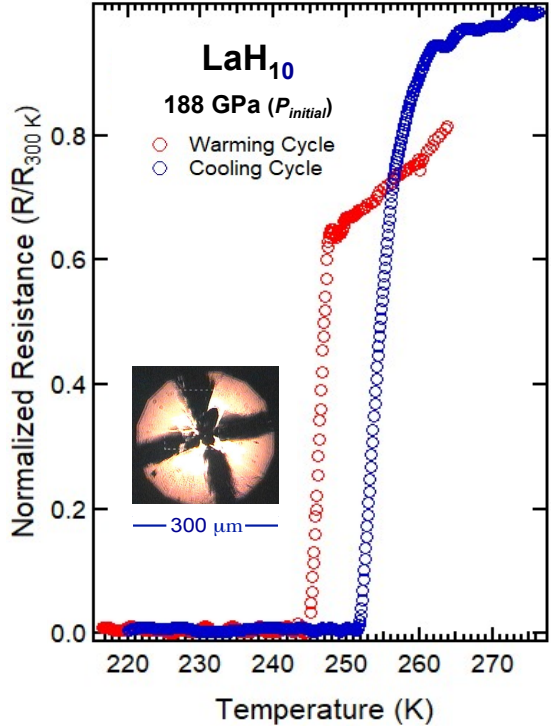

**Figure 4**. Normalized resistance of a $LaH_{10}$ sample characterized by x-ray diffraction and radiography, and measured with a four-probe technique. The inset shows an optical micrograph of the 35-$\mu$m sample and electrodes after synthesis.[31]

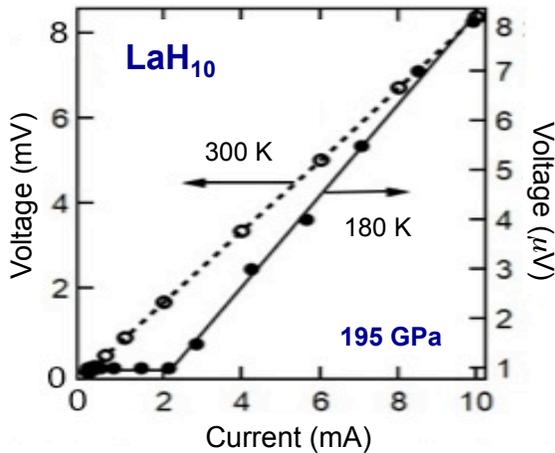

**Figure 5**. Current-voltage curves measured of a $LaH_{10}$ sample at 195 GPa at 300 K (mV scale) and 180 K ($\mu$V scale), well below $T_c$.[31]

Because of the smaller anvil type used in these experiments, a larger pressure gradient existed across the sample, which gives different phases in the sample. As a result, a series of phases were observed with $T_c$ ranging from 150 K to as high as 280 K. We have also explored the effect of synthesis at pressures below 170-180 GPa, and find lower maximum values for $T_c$, even upon tuning the transition as a function of pressure following synthesis. The results are also supported by preliminary Meissner effect measurements using a high-pressure magnetic susceptibility technique that we have used extensively to explore superconductivity at megabar pressures.[32,33]

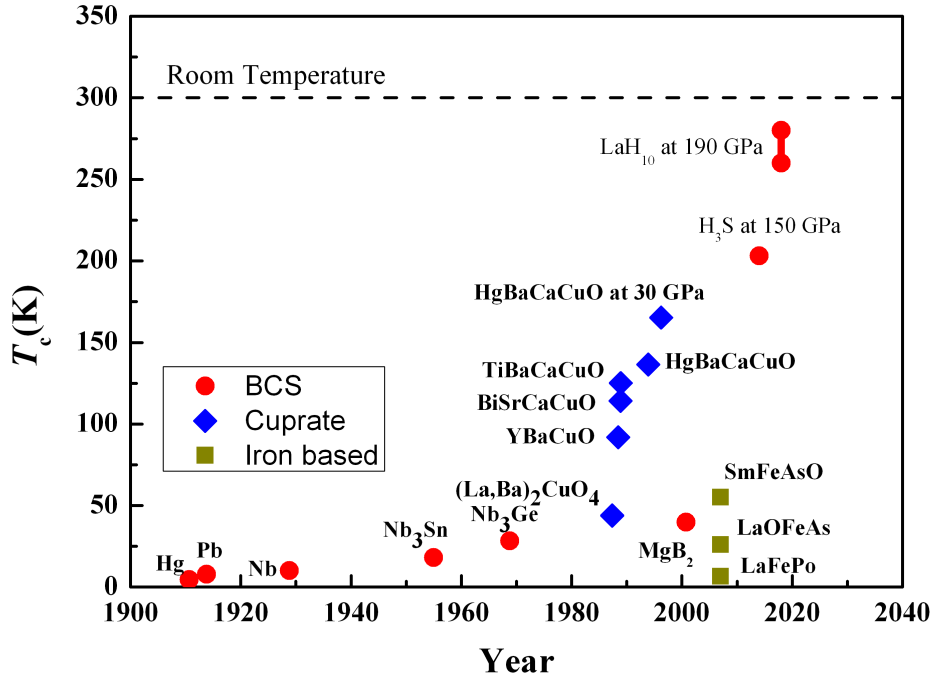

**Figure 6**. Observed $T_c$ over the years since the discovery of superconductivity in 1911.

**Discussion**

    Figure 6 shows the maximum $T_c$ over the years since the discovery of superconductivity in 1911. The trend for the conventional superconductors is shown as well as for the conventional superconductors with the hydrides and where we are now with the discovery of lanthanum superhydride. As indicated above, higher transition temperatures have been predicted, e.g., in the Y-H system,[26,27] but these are yet to be confirmed. Recently, the Mainz group has reported related studies of the La-H system at these pressures. The first report[34] contained electrical conductivity data that indicated a $T_c$ of 215 K for material synthesized at pressures below those required for cubic $LaH_{10}$, consistent with our theoretical and experimental results. A very recent paper[35] has confirmed the high $T_c$ cubic $LaH_{10}$ that we reported at this conference, though their synthesis pressures and maximum $T_c$ were slightly lower. That paper also reported evidence for a decrease in critical temperature with magnetic field as well as different synthesis procedures and starting materials, which provide further evidence that the phases represent the thermodynamic ground state of the material at these high pressures.

    These recent development are leading to a growing number of theoretical calculations. The results indicate strong electron-phonon coupling of the hydrogen atoms in the clathrate-like cage structures and hybridization of the La $4f$ and the H $1s$ orbitals. These calculations indicate that this coupling is associated with strong charge transfer from the La to the 32-atom hydrogen cage.[26,27,36,37] An alternative polaron model involving anionic La has also been proposed.[38] Yet to be explored in detail are nuclear quantum effects on the electron pairing mechanism in view of the evidence for the highly quantum character of this system even up to the critical temperature.[28] A related point has been made in the case of the $H_3S$ superconductor, where it has been proposed that the large zero-point motion of the hydrogen atoms in that system cause a breakdown of the Migdal approximation.[39] Another question is the extent to which the superhydrides of La and Y, and perhaps rare earth metals in general, may be unique in giving rise to the high $T_c$. Both La and

Y are adept at forming these 32 atom or larger cages. They are both *d*-elements and tend to form cubic hydrides with a closed shell of 32 hydrogen atoms, whereas addition of *f* electrons leads to other structures, which are predicted to have lower critical temperatures.[27,37]

The recent discoveries represent an example of the maturity of extreme conditions science, specifically, with respect to the century-old goal of realizing room-temperature superconductivity. The work also shows the importance of large-scale advanced radiation facilities (e.g., synchrotron x-ray sources) combined with high-pressure techniques in materials discovery. Hand in hand with advances in these experimental techniques is the increasing accuracy of theoretical and simulation methods that can guide experiments. Indeed, systematic prediction, synthesis, and validation will be important not only in understanding the mechanisms, but also for recovering materials exhibiting these properties for practical applications. One route is replacing the hydrogens with carbon to create analogous carbon-type clathrates such as $NaC_6$ (Fig. 7).[40] Further, we anticipate exploration of a much broader range of compositions, especially moving beyond simple binaries to ternaries and more complex chemical compositions. Judicious tuning of the structure and composition to optimize the electronic properties for maintaining very high $T_c$ superconductivity as well as stability (or metastability), for example within carbon frameworks, could lead to a new generation of superconductors that could operate at or near ambient pressures.

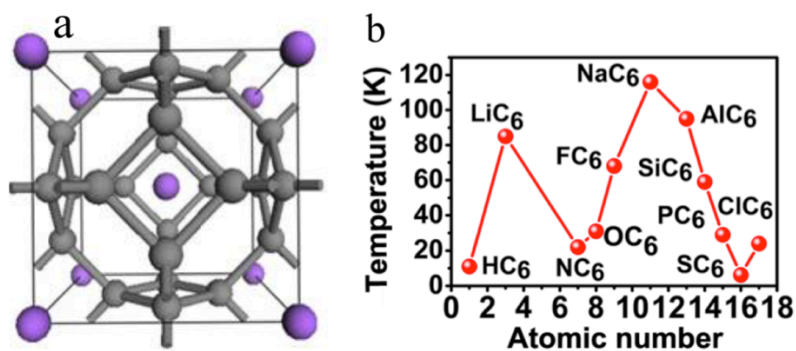

**Figure 7**. (a) Sodalite-type structure for predicted $XC_6$ carbides, showing the 24 atom cage also predicted for hydrides (e.g., $CaH_6$). (b) Predicted $T_c$ for different $XC_6$ compounds at ambient pressure. The calculations predict that $NaC_6$ is dynamically stable and therefore a potential superconductor at ambient pressure with an estimated $T_c$ of ~110 K.[40]

**Summary and Prospects**

The discovery of lanthanum superhydrides under pressure represents the newest phase in the road to room-temperature superconductivity. The phase identified as cubic $LaH_{10}$ superconducts with a $T_c$ above 260 K at pressures close to 200 GPa. Other superconducting phases are also observed, and the maximum $T_c$ depends on structure and synthesis pressures. The high $T_c$ has been subsequently confirmed in multiple experiments, including measurements of both electrical and diamagnetic properties. Available evidence suggests conventional superconductivity but further work is needed for detailed understanding of underlying mechanism, including a detailed treatment of the nuclear quantum effects. The creation of these and analogous very high $T_c$ superconductors that are stable (or metastable) under ambient conditions remains a grand challenge but may be possible with continued advances in synthetic methods and understanding of the structural and compositional control of this very high-temperature superconductivity. Future prospects include further exploration of the rich physics of hydrogen-rich systems at extreme conditions, continued development of 'materials by design' approaches in high-pressure science, and the potential for a new era of research in superconductivity opened by these recent developments.

*Acknowledgments:* We are grateful to Z. M. Geballe, I. I. Naumov, A. K. Mishra, Y. Meng, M. Baldini, R. Hoffmann, N. W. Ashcroft, V. V. Struzhkin, and S. A. Gramsch for their contributions to this work. This research was supported by NSF (DMR-1809783) and DOE/SC (Energy Frontier Research Center Program, DE-SC0001057). Instrumentation and facilities used were supported by the DOE/NNSA (DE-NA-0003858, CDAC; and DE-NA0001974, HPCAT), the Carnegie Institution of Washington, and DOE/SC (Contract DE-AC02-06CH11357 to Argonne National Laboratory).